# UML-F: A Modeling Language for Object-Oriented Frameworks

Marcus Fontoura[1], Wolfgang Pree[2], and Bernhard Rumpe[3]

[1] Department of Computer Science, Princeton University
35 Olden Street, Princeton, NJ 08544-2087, U.S.A
`mfontoura@acm.org`

[2] C. Doppler Lab for Software Research, University of Constance
D-78457 Constance, Germany
`pree@acm.org`

[3] Software and Systems Engineering, Munich University of Technology,
D-80290 Munich, Germany
`rumpe@acm.org`

**Abstract.** The paper presents the essential features of a new member of the UML language family that supports working with object-oriented frameworks. This UML extension, called UML-F, allows the explicit representation of framework variation points. The paper discusses some of the relevant aspects of UML-F, which is based on standard UML extension mechanisms. A case study shows how it can be used to assist framework development. A discussion of additional tools for automating framework implementation and instantiation rounds out the paper.

## 1  Introduction

Object-oriented (OO) frameworks and product line architectures have become popular in the software industry during the 1990s. Numerous frameworks have been developed in industry and academia for various domains, including graphical user interfaces (e.g. Java's Swing and other Java standard libraries, Microsoft's MFC), graph-based editors (HotDraw, Stingray's Objective Views), business applications (IBM's San Francisco), network servers (Java's Jeeves), just to mention a few. When combined with components, frameworks provide the most promising current technology supporting large-scale reuse [16].

A framework is a collection of several fully or partially implemented components with largely predefined cooperation patterns between them. A framework implements the software architecture for a family of applications with similar characteristics [26], which are derived by specialization through application-specific code. Hence, some of the framework components are designed to be replaceable. These components are called variation points or hot-spots [27] of the framework. An application based on such a framework not only reuses its source code, but more important, its architecture design. This amounts to a standardization of the application structure and allows a significant reduction of the size and complexity of the source code that has to be written by developers who adapt a framework.

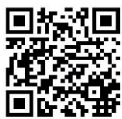

Recent standardization efforts of the Unified Modeling Language (UML) [32] offer a chance to harness UML as notational basis for framework development projects. UML is a multi-purpose language with many notational constructs, however, the current standard UML does not provide appropriate constructs to model frameworks. The constructs provided by standard UML are not enough to assist framework development, as will be discussed during the rest of this paper. There is no indication in UML design diagrams what are the variation points and what are their instantiation constraints. Fortunately, UML provides extension mechanisms that allow us to define appropriate labels and markings for the UML model elements.

This paper describes how to explicitly model framework variation points in UML diagrams to describe the allowed structure and behavior of variation points. For this purpose, a number of extensions of standard UML are introduced. The extensions have been defined mainly by applying the UML built-in extensibility mechanisms. These extensions form a basis for a new UML profile [7, 33, 35], especially useful for assisting framework development. This new profile is called UML-F.

The main goal of this paper is to introduce some key elements of UML-F and to demonstrate their usefulness. It would be beyond the scope of this paper to introduce the whole set of UML-F extensions. One of the main goals of defining UML-F was to try to use a small set of extensions that capture the semantics of the most common kinds of variation points in OO frameworks. In this way the designer can profit from his or hers previous experience with UML and learn just a few new constructs to deal with frameworks. This paper describes how the extensions have been defined allowing others extensions that deal with new kinds of variation points to be added to UML-F if needed. The current version of UML-F was refined based on the experiences of a number of projects [11]. These experiences have shown how UML-F can assist the framework development and instantiation activities to reduce development costs and at the same time increase the resulting quality of the delivered products. This paper presents a condensed version of a real-application case study to illustrate the benefits of UML-F and its supporting tools.

The rest of this paper is organized as follows: Section 2 outlines the UML extensions and discusses how they can be used to explicitly represent framework variation points. It also shows how the extensions allow for the development of supporting tools that can assist framework development and instantiation. Section 3 describes a case study of real application of UML-F, illustrating its benefits. Section 4 discusses some related work. Section 5 concludes the paper and sketches our future research directions.

## 2   The Proposed UML Extensions

This section introduces UML-F through an example. It summarizes the new extensions and presents a general description of their semantics. It also presents a description of the UML extensibility mechanisms and how they have been applied in the definition of UML-F. A description of tools that use UML-F design descriptions to automate framework development and instantiation is also presented.

## 2.1 Motivating Example

Figure 1 shows a student subsystem of a web-based education framework [12] in plain UML, where (a) represents a static view of the system (UML class diagram) and (b) provides a dynamic view (UML-like sequence diagram). The dynamic view illustrates the interaction between an instance of each of the two classes.

The *showCourse()* method is the one responsible for controlling the application flow: it calls *selectCourse()*, which allows the student to select the desired course, *tipOfTheDay()*, which shows a start-up tip, and finally *showContent()* to present the content of the selected course.

Method *selectCourse()* is the one responsible for selecting the course the student wants to attend. It is a variation point since it can have different implementations in different web-based applications created within the framework. Different examples of common course selection mechanisms include: requiring a student login, showing the entire list of available courses or just the ones related to the student major, showing a course preview, and so on. There are numerous possibilities that depend on the framework use.

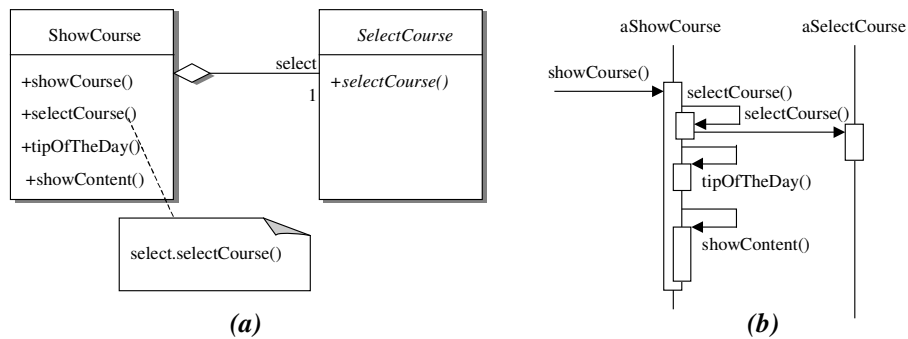

**Figure 1.** UML representation of a framework web-based framework.

Figure 1 shows *selectCourse()* as an abstract method of an abstract class *SelectCourse*. During framework instantiation, the framework users would have to create subclasses of *SelectCourse* and then provide a concrete implementation of the *selectCourse()* method. The problem with this representation is that there is no indication that *selectCourse()* is a variation point in the design diagrams. There is also no indication of how it should be instantiated. Although the name of the abstract method *selectCourse()* is italicized this notation is not an indication of a variation point, rather it indicates an abstract method which does not necessarily have to be a variation point.

Method *tipOfTheDay()* is also a framework variation point. The reason is that some applications created from the framework might want to show tips while others will not do so. The framework should provide only the methods and information that are useful for all the possible instantiated applications and the extra functionality should be provided only in framework instances. Although this may seem a strong statement, it is the ideal situation. The inclusion of methods like *tipOfTheDay()* could lead to a complex interface for *ShowCourse*, with many methods that would not be needed by several framework instances. A good design principle in designing a framework its to try to keep it simple; extra functionality can always be placed in component libraries.

The *Actor* class hierarchy is used to let new types of actors be defined depending on the requirements of a given framework instance. The default actor types are students, teachers, and administrators, however, new types may be needed such as librarians, and secretaries. This means that applications created from the framework always have at least three kinds of actors, students, teachers, and administrators, but several other actor types may be defined depending on the application specific requirements. This design structure is presented in Figure 2.

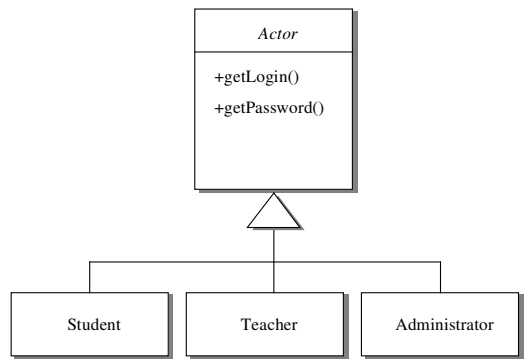

**Figure 2.** Actor hierarchy.

The *Actor* class hierarchy also represents a variation point, since it allows the definition of new classes to fulfill the application specific requirements. However, this is not properly indicated in the UML diagram presented in Figure 2. The framework developer should be able to indicate the variation points in class hierarchies to facilitate the job of the framework user during the instantiation process. Fortunately, UML provides a constraint called *Incomplete* in its standard set of constraints. *Incomplete* indicates that new classes may be added to a given generalization relationship and was adopted as part of UML-F, as will be described in subsection 2.3.

### 2.2 UML Extensibility Mechanisms

UML provides three language extension mechanisms: stereotypes, tagged values, and constraints. Stereotypes allow the definition of extensions to the UML vocabulary, denoted by *«stereotype-name»*. Each model element (e.g. a class or a relationship) can have a stereotype attached. In this case, its meaning is specialized in a particular way suited for the target architecture or application domain. A number of possible uses of stereotypes have been classified in [2], but stereotypes are still a rather new concept and still subject of ongoing research [7].

Tagged values are used to extend the properties of a modeling element with a certain kind of information. For example, a version number or certain tool specific information may be attached to a modeling element. A tagged value is basically a pair consisting of a name (the tag) and the associated value, written as "*{tag=value}*". Both tag and value are usually strings only, although the value may have a special interpretation, such as numbers or the Boolean values. In case of tags with Boolean values, UML 1.3 allows us to write "*{tag}*" as shortcut for "*{tag=TRUE}*". This leads to the fancy situation that occasionally concepts a stereotype, e.g. *«extensible»*, and a tag, e.g. *{extensible}*, could be used for the same purpose. Since model elements can only have one stereotype, but an unlimited number of tagged values, it is often better to use tagged values in this kind of situation. They provide more flexibility, e.g. freeing us of defining a new stereotype for each combination of tags that may be attached to a model element.

In addition to the mentioned two UML extension mechanisms, there exist constraints. Constraints may be used to detail how a UML element may be treated. However, like the other two, constraints have a rather weak semantics and therefore can be used (and misused) in a powerful way. Constraints are today usually given informally, or by a buzzword only. The *{incomplete}* constraint (Figure 3) could also be defined as tagged value.

We expect that this mismatch among the extensibility mechanisms be improved in future UML versions. D'Souza, Sane, and Birchenough suggest that all three kinds of extensions should be stereotypes [7]. We argue in favor of this unification, but we will retain the flexibility of tags and therefore will use tagged values for all purposes.

**2.3 UML-F Extensions**

This subsection introduces UML-F illustrating its application to model the web-based education framework [12]. Figure 3 models part of the framework representing and classifying the variation points explicitly. The variation points are modeled by a number of tagged values with values of Boolean type to extend the UML class definitions.

In this example the method *selectCourse()* is marked with the tagged value *{variable}* to indicate that its implementation may vary depending on the framework instantiation. The tagged value *{variable}* has the purpose to show the framework user that *selectCourse()* must be implemented with application specific behavior for each framework instance. Methods marked with *{variable}* are referred to as *variable methods*.

In contrast to the previous tagged value, *{extensible}* is applied to classes. In this example *{extensible}* is attached to the *ShowCourse* class, indicating that its interface may be extended during the framework instantiation by adding new functionality, like methods such as *tipOfTheDay()*. Please note that extension is optional, but not a must.

An important point here is that the diagram shown in Figure 3 is a result of a design activity, and therefore may implemented in several different ways. The fact that a class is marked as *{extensible}* tells us that its implementation will have to allow for the extension of its interface, since a given framework instance may want to do so. However, it does not mean that the new methods have to be added directly to the class. The same holds for variable methods: the changes may be defined without changing the method directly, but by the addition of new classes that provide appropriate implementations for the method. Section 3 discusses some implementation techniques that may be applied to model variable methods and extensible classes.

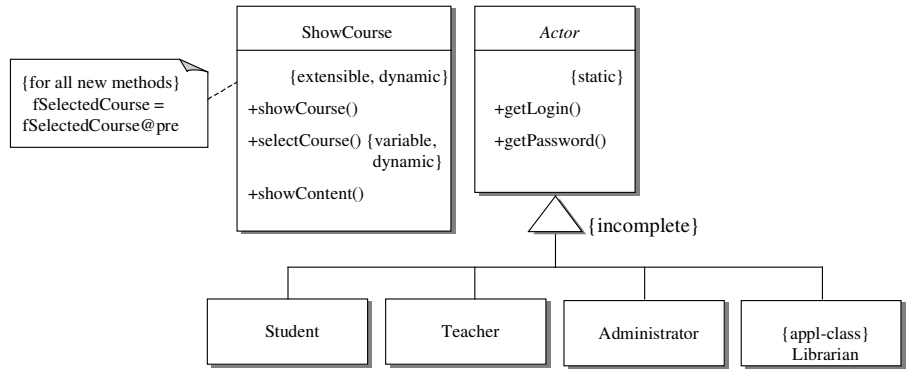

**Figure 3.** UML-F extended class diagram.

Figure 3 uses the tag *{incomplete}* to indicate a third kind of variation point: an *extensible interface*. *{Incomplete}* is applied to a generalization relationship, allowing new subclasses to be defined by framework instances. In this example it indicates that new subclasses of *Actor* may be provided to fulfill the requirements of applications created from the framework. Please note that *{incomplete}* is already provided by the UML as a constraint, with exactly the same meaning used here.

The tag *{appl-class}* is used to indicate a placeholder in the framework structure where *application specific classes* may be or have already been added. It complements the definition of extensible interfaces: the generalization relationship between an extensible interface and an application class is always *{incomplete}*. Class *Librarian* is an example of an *application class*. The *{incomplete}* tag allows the framework user to create as much application classes from a given extensible interface during framework instantiation as needed. In contrast to the other two kinds of variation points, extensible interfaces have a direct mapping from design to implementation since current OO programming languages provide constructs for modeling generalization relationships directly.

Two other Boolean value tags, called *{dynamic}* and *{static}*, complement the variation point definition by indicating whether runtime instantiation is required. Each variation point can be marked either by the *{dynamic}* or by the *{static}* tag (but not both). Variable methods are instantiated by providing the method implementation. Extensible classes are instantiated by the addition of new methods. Extensible interfaces are instantiated by the creation of new application classes. Interpreted languages, such as Smalltalk and CLOS, give full support for runtime, or *{dynamic}*, instantiation. Java offers dynamic class loading and reflection that also can be used to allow dynamic instantiation of variation points. In the example shown in Figure 3 the tag *{dynamic}* is used because it is a user requirement to have dynamic reconfiguration for the variation points that deal with course exhibition. The tag *{static}* is used for the *Actor* extensible interface since new actor types do not need to be defined during runtime. The tag *{dynamic}* implies that the implementation has support for runtime instantiation for the marked element. However, such a runtime instantiation must not necessarily happen.

The note attached to the *ShowCourse* extensible class is an OCL [25, 33, 35] formula that defines that the class attribute *fSelectedCourse* shall not be changed by any of the new methods that may be added to the *ShowCourse* extensible class during framework instantiation. This kind of restrictions over variation points is called instantiation restrictions. To be able to describe certain OCL constraints for methods that have neither been introduced nor named yet the tag *{for all new methods}* is used, indicating that this constraint is to hold for all new methods. This kind of tag strongly enhances the power of description of the design language, as it allows us to talk about methods that have not even been named yet.

Although it is beyond the scope of this paper, Figure 4 shows a sequence diagram that can be used to limit the possible behavior of a variation point. The sequence diagram shows the main interaction pattern for a student selecting a course. As it may be decided by actual implementation, it is optional whether the student has to log in before he selects a course or whether the data is validated. This kind of option can be shown in sequence diagram by using *{optional}* tag, which indicates interactions that are not mandatory. In the area of sequence diagrams, there are many more possibilities to apply tags of this kind for similar purposes, such as determining alternatives, avoidance of interleaving, and so on. We expect useful and systematic sets of tags for sequence diagrams to come up in the near future. Figure 4 tells us that a concrete method that instantiates *selectCourse()* must have the following behavior:

1. It may display a login web page;

2. It must show a web page for the selection of the desired course;

3. It may validate the data by checking if the login is valid, and whether the student is assigned to the course or not. This step is optional since there can be courses that do not require student identification;

The extended class diagrams and the sequence diagrams complement each other providing a rather useful specification of variation points and their instantiation restrictions. It is important that framework developers provide documentation that describes what parts of the system should be adapted to create a valid framework instances. It is quite cumbersome that framework users today often need to browse the framework code, which generally has complex and large class hierarchies to try to identify the variation points. The diagrams and diagram extensions introduced in this example address this problem. Section 3 will further discuss these ideas, showing how UML-F can assist framework implementation and instantiation.

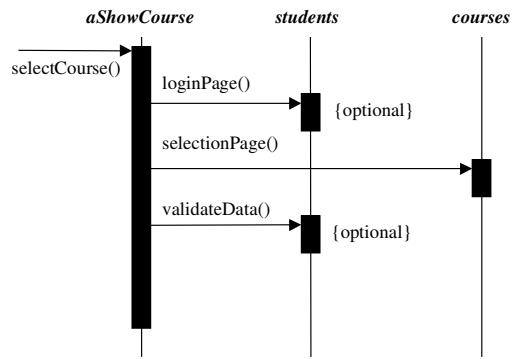

**Figure 4.** Sequence diagram for *selectCourse()*.

### 2.4 Language Description

Once the extensions are defined it is crucial to specify their exact meaning. As a sidenote, it is important to mention that in most languages (such as natural language, like English), new vocabulary is explained through a definition using existing vocabulary. This even holds for programming languages, like Java, where new classes and methods are defined using existing classes, methods, and basic constructs. Unfortunately, UML 1.3 and high likely also UML 1.4 does not provide a clear path for defining the precise semantics of new stereotypes, tagged values, and constraints. Therefore, this section describes the meanings of our newly introduced elements mainly informally. A formal approach to characterize a variant of these elements based on set theory is presented in [11]. However, this formal definition of UML-F is not presented here since its usefulness for the communication purposes is limited [33].

This paper demonstrates how UML-F deals with three kinds of variation points: *variable methods*, *extensible classes*, and *extensible interfaces*. Variable methods are methods that have a well-defined signature, but whose implementation varies for each instantiated application. In the example *selectCourse()* is a variable method. Extensible classes are classes that may have their interfaces extended during the framework instantiation. *ShowCourse*, for example, may require the addition of new methods (like *tipOfTheDay()*) for each different application. Extensible interfaces are interfaces or abstract classes that allow the creation of concrete subclasses during the framework instantiation. The instantiation of this last kind of variation point takes place through the creation of new classes, called *application classes*, which exist only in framework instances.

It should be clear that these three kinds of variation points have different purposes: in variable methods the method implementation varies, in extensible classes the class interface varies, finally, in extensible interfaces the types in the system vary (new application classes may be provided). All three kinds may either be static (do not require runtime instantiation) or dynamic (require runtime instantiation).

There are other kinds of variation points in framework design, such as variation in structure (attribute types for example). Coplien describes several kinds of variability problems in his multi-paradigm design work [6]. They integrate well into UML-F using similar principles to the ones described in this paper. To avoid the explosion of the number of extensions and to keep the presented part of UML-F feasible this paper focus on the most important kinds of variation points.

UML diagrams are extended by the tags *{variable}, {extensible}, {incomplete}, {appl-class}, {static},* and *{dynamic}*. The first two represent variable methods and extensible classes, respectively. *{Static}* and *{dynamic}* are used to classify them regarding to their runtime requirements. The *{incomplete}* tag (in UML 1.3 known as constraint) has been adapted to identify extensible interfaces. The keywords *{extensible}, {variable},* and *{incomplete},* indicate what are the variation points and their exact meaning. The *{appl-class}* stereotype indicates placeholders for classes that are part of instantiated applications only.

OCL specifications [25, 33, 35] may be written on notes as in standard UML, however, they have an enhanced meaning if the notes are attached to variation points. In the case of variable methods, it means that all method implementations that may be defined during instantiation should follow the specification. If an OCL constraint is attached to an extensible class, the special tag *{for all new methods}* is useful to describe the behavior of methods that do not even have a name yet. This tag indicates that the constraint applies to all methods that might be added during instantiation. Similarly, if attached to an extensible interface, the OCL constraint applies to all methods that can be overridden or added to each application class.

Let us also mention the tag *{optional}*. Here, it extends sequence diagrams to indicate that certain interaction patterns are not obliged to occur. These sequence diagrams have proven useful to be applied to all kinds of variation points. Generally, they are used to describe a *pattern behavior* that should be followed by the variation point instances, as shown in Figure 4. OCL specifications, on the other hand, are generally used to specify invariants that should be satisfied by the variation point instances, as shown in Figure 3. Thus, sequence diagrams and OCL constraints complement each other in constraining the possible instantiations of variation points, and may therefore be used together.

Table 1 summarizes the new UML-F elements and informally defines their semantics.

### 2.5 Tool Support

This subsection shows how tools that benefit from the UML-F design diagrams may be defined to assist both framework development and instantiation. The tools suggested here have a prototypical implementation using PROLOG. However, many currently available UML case tools give support reasoning about tagged values and could be adapted to work with UML-F. This subsection gives information to allow the customization of UML case tools for working with OO frameworks.

**Table 1.** Summary of the new elements and their meanings

| Name of extension | Type of extension | Applies to notational element of UML | Description |
|---|---|---|---|
| *{appl-class}* | Boolean Tag | Class | Classes that exist only in framework instances. New application classes may be defined during the framework instantiation. |
| *{variable}* | Boolean Tag | Method | The method must be implemented during the framework instantiation. |
| *{extensible}* | Boolean Tag | Class | The class interface depends on the framework instantiation: new methods may be defined to extend the class functionality. |
| *{static}* | Boolean Tag | Extensible Interface, Variable Method, and Extensible Class. | The variation point does not require runtime instantiation. The missing information must be provided at compile time. |
| *{dynamic}* | Boolean Tag | Extensible Interface, Variable Method, and Extensible Class. | The variation point requires runtime instantiation. The missing information may be provided only during runtime. |
| *{incomplete}* | Boolean Tag | Generalization and Realization | New subclasses may be added in this generalization or realization relationship. |
| *{for all new methods}* | Boolean Tag | OCL Constraint | Indicates that the OCL constraint is meant to hold for all newly introduced methods. |
| *{optional}* | Boolean Tag | Events | Indicates that a given event is optional. It is useful for specifying a template behavior that should be followed by the instantited variation point. |

**Assisting Framework Development.** Standard OO design languages do not provide constructs for representing flexibility and variability requirements. UML-F addresses this problem representing variation points as first-class citizens thus making the framework intentions more explicit. The new language elements are not concerned with how to implement the variability and extensibility aspects of the framework, but focus on representation at design level. Consequently, the diagrams are more abstract (and more concise) than standard OO diagrams. Unfortunately some of the new design elements cannot be directly mapped into existing OO programming languages.

Extensible interfaces can be directly implemented through standard inheritance. Although dynamic extensible interfaces are not supported in compiled languages such as C++, they may be simulated through dynamic linking (Microsoft Windows DLLs, for example). Variable methods and extensible classes, on the other hand, cannot be directly implemented, since standard OO programming languages do not provide appropriate constructs to model them.

To bridge this design-implementation gap, several techniques may be used. Design patterns are a possible solution, since several patterns provide solutions for flexibility and extensibility problems and are based only on extensible interfaces. Thus, design patterns may be used to transform variable methods and extensible classes into extensible interface variation points. Figure 5 illustrates the use of the Strategy design pattern [15] to implement this mapping. Classes *ShowCourse* and *SelectStrategy* are identified with the tags *{separation, template}* and *{separation, hook}* to indicate the roles they play in the pattern. Strategy is based on the Separation meta-pattern [28], in which a template class is responsible for invoking the variable method in the hook class. The use of tags that indicate meta-pattern roles complement the UML-F description for variation points implemented by design patterns, further clarifying the design. A similar solution for identifying design diagrams with pattern roles is described in [30].

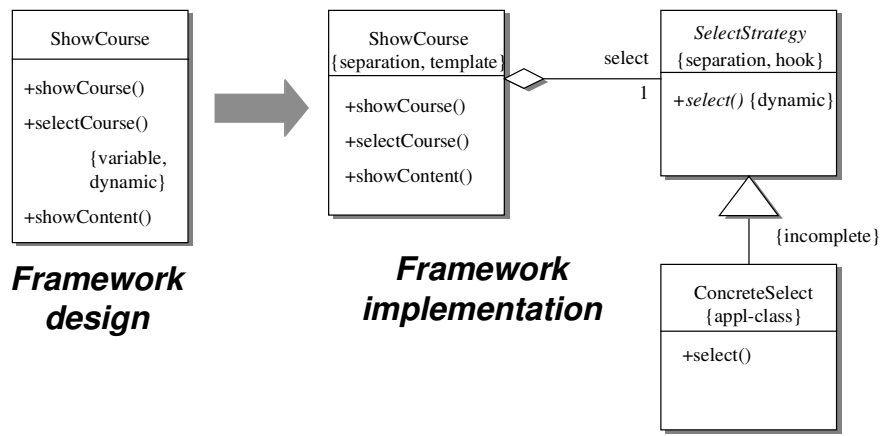

**Figure 5.** Transforming variable methods into extension interface variation points.

The transformations used to map variable methods and extensible classes into implementation level constructs must be behavior-preserving, since the system functionality is independent of the implementation technique used to model the variation points.

A *code generation* tool can be used to automate design to implementation transformations. It is responsible for mapping the new design elements of UML-F into appropriate implementation level structures. More specifically, it is responsible for eliminating the variable methods and extensible classes from the design. This mapping is based on meta-artifacts that describe the transformations. These meta-artifacts are called implementation models. It is an imperative to allow the definition of new implementation models for variation points, so that different styles of translation are possible.

The transformation illustrated in Figure 5 is an example of a mapping supported by the code generation tool. The implementation model that supports this transformation describes how dynamic variable methods are modeled by the Strategy design pattern. Figure 6 illustrates the code for this implementation model, which searches for all variable methods in the design diagrams and applies Strategy to them.

The implementation transformations (illustrated in Figure 6) preserve the design structure described in *Project* and create *NewProject* to store the generated framework. All the design elements that are not transformed, the kernel elements and the extensible interfaces, are copied from *Project* to *NewProject*. The variable methods and extensible classes are transformed in the way described by the selected implementation model.

```
applyStrategy(Project, NewProject) :-          ← Searches for
        [...]                                     variable
        forall(variableMethod(Project, Class, Method, _), methods
        strategy(Project, NewProject, Class, Method)),
        [...]
                                               ← Uses strategy
strategy(Project, NewProject, Class, Method) :-  to model them
        concat(Method, 'Strategy', NewClass),
        createExtensibleInterface(NewProject, NewClass, dynamic),
        createMethod(NewProject, NewClass, Method, public, none, abstract),
        createAggregation(NewProject, Class, NewClass, strategy),
        [...]
```

**Figure 6.** Strategy implementation model.

Each valid implementation model artifact has to define at least four transformations: (static and dynamic) variable methods and (static and dynamic) extensible classes. Examples of implementation models that have been successfully used to assist framework implementation include different combinations of design patterns, meta-programming [21], aspect-oriented programming (AOP) [20], and subject-oriented programming (SOP) [17], as described in [11]. The case study section also describes some other mappings.

The selection of the most appropriate technique to be used model each variation point is a creative task and cannot be completely automated. However, UML-F diagrams and the set of implementation models available for each kind of variation point may help the framework designer to narrow his or hers search for appropriate implementations. Moreover, the code generation tool automatically applies the transformation once the implementation model has been selected, making the mapping from design to implementation less error prone.

Some UML case tools, such as Rational Rose (http://www.rational.com), allow the customization of how code is generated from the design diagrams. Therefore, it is possible to specify how code should be generated for the new UML-F elements.

**Assisting Framework Instantiation.** During the framework instantiation, application classes must be provided to complete the definition of the extensible interface variation points (at this point this is the only kind of variation points in the system, given that the other two have already been eliminated during implementation). Figure 7 illustrates a framework instantiation. After the instantiation all extensible interfaces disappear from the design, since the *{incomplete}* generalizations become "complete." In this example the variation point was instantiated by just one concrete application class, *SimpleSelect,* which is marked by the *{c-hook}* tag to indicate that it plays the role of a concrete hook. In a general case, however, several application classes may be provided for each extensible interface.

An instantiation tool can be used to assist the application developer to create applications from the framework. The tool knows what are the exact procedures to instantiate extensible interfaces: it has to create a new subclass, ask for the implementation of each of the interface methods, and ask for the definition (signature and implementation) for each new method that might be added, if any. The tool prompts the application developer about all the required information to complete the missing information for each variation point in the framework structure.

Note that the tags that indicate the meta pattern roles are useful just for enhancing the design understating, and are not processed by the implementation and instantiation tools.

Depending on the implementation model selected, different instantiation tasks may be required for the same variation point, as will be illustrated in Section 3. UML-F descriptions can be seen as structured cookbooks [22] that precisely inform were application specific code should be added. The instantiation tool is a wizard that assists the execution of these cookbooks. Once again the code generation part of standard UML case tools may be adapted to mark the points in which code should be added by using the information provided by the extensible interface tags.

## 3  Case Study

This section details the implementation and instantiation of the web-education framework modeled in Figure 3. It starts from the UML-F specification, derives the final framework implementation, and shows how it may be instantiated. The benefits of UML-F and its supporting tools are discussed throughout the example.

### 3.1 Framework Implementation

Let us consider that the only variation points of the framework are the ones presented in Figure 3. Since all the variation points have been identified and marked in the UML-F design diagrams, the next step is to provide implementation solutions to model them. As discussed before, extensible interfaces and the framework kernel (modeled only by standard UML constructs) have straightforward mappings into OO programming languages. Therefore the framework designer focus during the implementation phase should be on how to model variable methods and extensible classes. In this example two variation points have to be examined: the *selectCourse()* variable method and the *ShowCourse* extensible class.

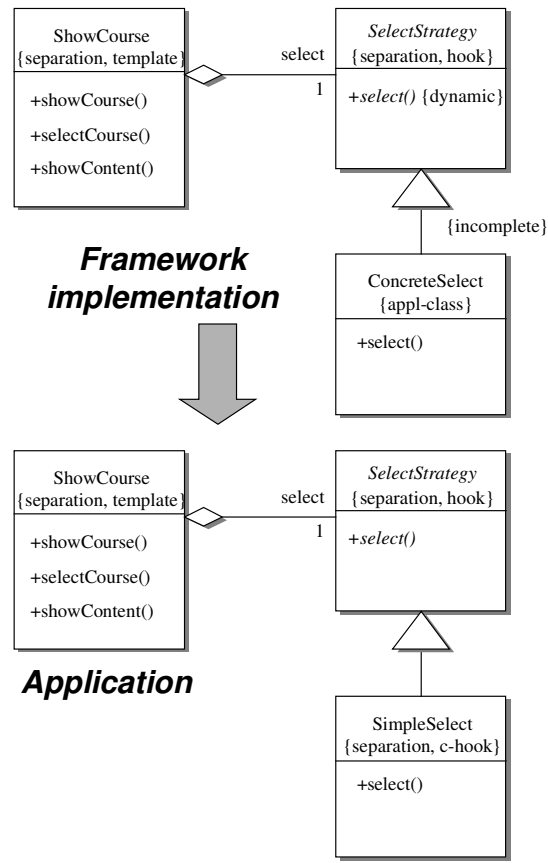

**Figure 7.** Instantiation example.

The designer has to select an appropriate technique based on his or hers experience. If a supporting tool with a set of implementation models is available, the analysis of these models may facilitate this task. One of the models available in the code generation tool is the use of the Strategy design pattern [15] to implement dynamic variable methods and a slightly changed version of the Separation meta-pattern [28], which allows the invocation zero or more hook methods, to implement dynamic extensible classes. Since the transformations are automatically applied by the tool let us try this solution and see what happens. The resulting design is shown in Figure 8.

This solution worked quite well. The solution for extending the *ShowCourse* interface allows the addition of new methods without directly changing the class interface. It allows an instance application to define zero or more methods that will be invoked before the actual content of the course is displayed, and that is the expected behavior. An important point to make is that the instantiation restriction specified by the OCL constraint in Figure 3 is automatically assured by this solution, since the new methods do not have access to the *fSelectCourse* attribute that is private to *ShowCourse*.

In the case of *selectCourse()*, however, the Strategy solution does not guarantee that the behavior specified by the sequence diagram in Figure 4 will be followed. Strategy is a white-box pattern since it allows the definition of any behavior for the hook method. The verification of this kind of instantiation restrictions is not an easy task (and is generally an undecidable one), however there are some implementation solutions that may be more restrictive, or more black-box.

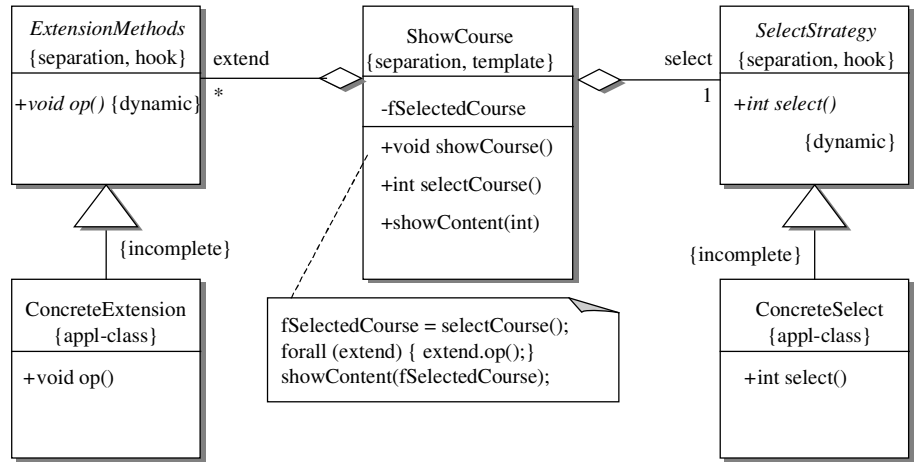

**Figure 8.** A pattern-based implementation.

A solution that might be more appropriate for *selectCourse()* is the definition of a meta-object protocol (MOP) [18]. MOPs allow meta-level concepts to be dynamically defined in terms of base-level ones. Thus, the use of MOP may be a good alternative since it is a more restrictive solution than the Strategy pattern: the possible instantiations are just the ones defined by the protocol. Figure 9 illustrates the use of MOP for this example. Whenever instances of the *SelectMOP* class are created a set of Boolean parameters that complete the variation point behavior have to be provided: *login* (TRUE if login is required), *major* (TRUE if a student can attend only the courses related to his or hers major), and *validate* (TRUE if it is required that the student have to be assigned to be able to attend the course). The combination of these parameters provides all the possible instantiations allowed by the MOP. Note that this solution is much more restrictive than the Strategy solution, but it has the advantage that it always preserves the instantiation restrictions specified in the sequence diagram.

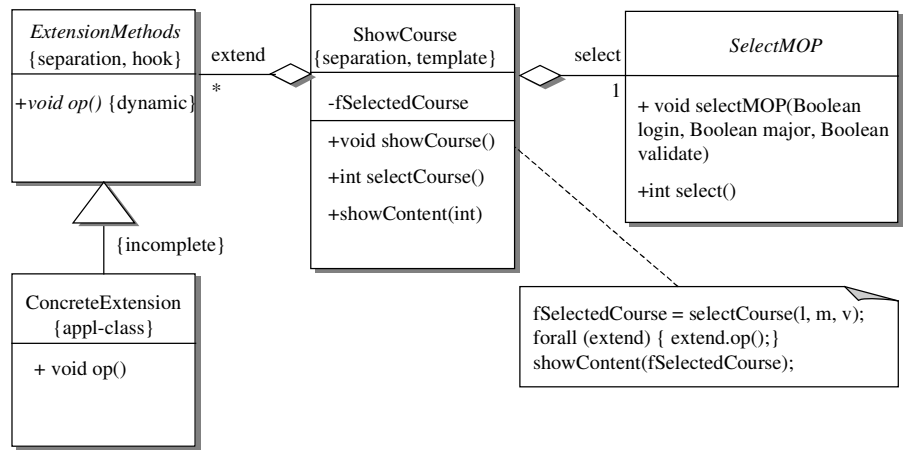

**Figure 9.** Using MOP to model selectCourse().

The implementation of MOPs cannot be automated by the code generation tool, since each MOP is specific for a given variation point. However, the UML-F instantiation restrictions provide a good documentation that can be used by the MOP developers. In this example the parameters *login* and *validate* can be directly derived from Figure 4. In general MOPs may require objets more complex than Boolean ones as parameters and reflection may be required in their implementation.

Note that the runtime constraints *{static}* and *{dynamic}* play a crucial role during framework development. In this example, if the variation points were defined as *{static}* a much simpler design solution based on the Unification meta-pattern [28] could be used for both cases. In Unification-based patterns the template and hook methods belong to the same class, leading to a less flexible but simpler design solution.

### 3.2 Framework Instantiation

During instantiation the variation points missing information have to be filled with application specific code. Since the variable methods and extensible classes have been eliminated during implementation, only extensible classes are left to be instantiated by the application developers.

Tools such as the instantiation tool may facilitate this task by identifying all the points in which code has to be written. However, even if no tools are available, the UML-F diagrams make this task very straightforward since all the extensible interfaces and their corresponding instantiation restrictions are marked in the diagrams.

Figure 10 shows an example of application created from the framework defined in Figure 8. Application classes are provided to complete the definition of the two variation points. Note that if the MOP solution had been adopted the *selectCourse()* variation point would not require new application classes, since MOPs are completely instantiated during runtime by parametrization. This illustrates that different implementation models applied to the same variation point may demand different instantiation procedures.

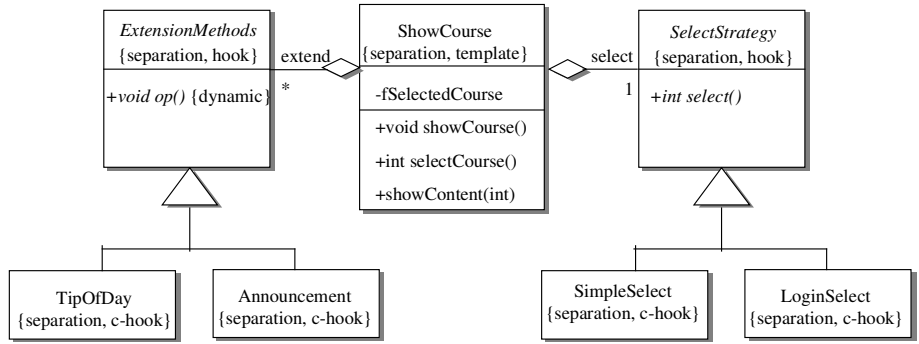

**Figure 10.** An application created from the framework.

## 4   Related Work

This section describes some of the current design techniques used to model frameworks, and relates them to UML-F. It shows that currently proposed constructs used to represent framework variation points have not adequately met our expectations.

Early OO design methods, like OMT [31], as well as the current UML 1.3, provide a number of diagrams for structure, behavior, and interaction. Different OO design notations include different artifacts, such as the representation of object responsibilities as CRC cards [1, 37]. However none of these artifacts has explicit support for the representation of the variation points of a framework.

UML represents design patterns as collaborations (or mechanisms) and provides a way of modeling framework adaptation through the binding stereotype [32]. However, framework instantiation usually is more complex than simply assigning concrete classes to roles: variation points might have interdependencies, might be optional, and so on. Catalysis uses the UML notation and proposes a design method based on frameworks and components [8]. Frameworks are treated in Catalysis as collaborations that allow substitution. However, as discussed in the paper, OO application frameworks may require different instantiation mechanisms. Therefore, Catalysis and standard UML only partly address the problems identified in this paper due to a lack of support for explicit marking variation points and their semantics.

Design patterns [4, 15, 36] are usually described using standard OO diagrams. Since various design patterns provide solutions to variability and extensibility problems [15] they define a common vocabulary to talk about these concepts [36] and may enhance the understanding of framework designs. Sometimes design pattern names are used as part of the class names allowing the framework user to identify variation points through the used names. However, in a typical framework design a single variation point class can participate in various design patterns. Then the approach of using design pattern names as class names becomes obfuscated. One possible solution for this problem is the use of role-based modeling technique, as shown in [30].

Meta-level programming [21], which can be seen as an architectural pattern [4], provides a good design solution for allowing runtime system reconfiguration. Therefore, the use of meta-level programming is a useful technique for modeling variation points that require runtime instantiation, and (with appropriate conventions) it may facilitate the identification of variation points in the framework structure. The case study shown in section 3 has shown that both design patterns and meta-level programming can be used in conjunction with UML-F, during the implementation of variation points.

The use of role diagrams to represent object collaboration is a promising field in OO design research [5]. Riehle and Gross propose an extension of the OOram methodology [29] to facilitate framework design and documentation [30]. His work proposes a solution for an explicit division of the design, highlighting the interaction of the framework with its clients. The use of roles does simplify the modeling of patterns that require several object collaborations and provides a solution for documenting classes that participate in several design patterns. However, no distinction is made between the kernel and variation point elements. This problem is handled using design patterns: if the framework user knows what patterns were used to model each of the variation points he or she can have an intuition on how the framework should be instantiated. On the other hand, if the pattern selections are not explicitly represented, the identification of the variation points becomes again difficult. Another disadvantage of this approach is the solution for modeling unforeseen extensions proposed in [30], which may lead to a very tangled design. Although it can be a good solution it should have a more concise representation at design level. This paper has shown how to use roles to complement the description of variation points implemented by design patterns.

Contracts [18, 19] and adaptable plug-and-play components (APPCs) [24] provide linguistic constructs for implementing collaboration-based (or role-based) diagrams in a straightforward manner. They may be used to implement variation points since they represent instantiation as first-class citizens. However, these concepts are still quite new and their use for implementing frameworks needs further investigation. Also Lieberherr and the researchers of the Demeter Project [24] have developed a set of concepts and tools to help and evaluate OO design that can be used to enhance framework development.

The Hook tool [13, 14] uses an extended version of UML in which the variation point classes are represented in gray. This differentiation between kernel and variation points helps framework design and instantiation, but it does not solve the problem completely. Framework designers still have to provide the solutions for modeling each variation point without any tool support. A good point of this approach is that instantiation constraints are treated as first-class citizens in the definition of hooks.

Several design pattern tools [3, 9, 10, 23] have been proposed to facilitate the definition of design patterns, to allow the incorporation of patterns into specific projects, to instantiate design descriptions, and to generate code. However, they leave the selection of the most adequate pattern to model each variation point in the hands of the framework designer. Although this is obviously a creative task, if variation points are modeled during design tools that assist the systematization of the selection of the best modeling technique for each variation point may be constructed, simplifying the job of the framework designer.

## 5   Conclusions and Future Work

The standardization of the UML modeling language makes it attractive as a design notation for modeling OO frameworks. This paper shows that UML today lacks constructs to explicitly represent and classify framework variation points and their instantiation restrictions. The proposed extensions to the UML 1.3 design language address this problem representing variation points through appropriate markings. They make the framework design more explicit and therefore easier to understand and instantiate. The extensions have been defined by applying the UML extension mechanisms.

Although the extensions described in this paper have been used to model frameworks successfully [11], they are neither complete nor the only ones that may be applied to framework development. This paper discusses how to improve UML-F to provide additional extensions and a systematic approach to apply these extensions to different kinds of UML diagrams. Furthermore, it is of interest to understand that relationship of UML-F with similar kinds of variability problems, such as presented in [6].

The new UML-F elements are not concerned with how to implement the variability and extensibility aspects of the framework, but just with how to appropriately represent them at the design level. Furthermore, through use of this kind of extensions it is more likely that the framework user will not have to go into the detailed internals of a framework, being able to use it in a more black-box manner. Consequently, the diagrams give us a more abstract and concise representation of a framework, when compared to standard OOADM diagrams.

The most important claims of this paper is that frameworks should be modeled through appropriate design constructs that allow the representation of variation points and their intended behavior. The extended class diagrams and sequence diagrams facilitate the definition of adequate documentation, which may be used to assist the framework developer in modeling the variation points and the framework user in identifying these points during instantiation.

The extensions allow for the definition of supporting tools that may partially automate the development and instantiation activities. Appropriate tool assistance should also lead to a better time-to-market, reduced software costs, and higher software quality.